\documentclass[conference]{IEEEtran}
\IEEEoverridecommandlockouts
\usepackage{cite}
\usepackage{amssymb,amsfonts}
\usepackage[tbtags]{amsmath}
\usepackage{algorithmic}
\usepackage{graphicx}
\usepackage{textcomp}
\usepackage{xcolor}
\usepackage{float}
\usepackage{cite}
\usepackage{bm}
\usepackage{epstopdf}
\usepackage{url}
\usepackage{lipsum}
\usepackage{hyperref}
\usepackage{array}
\usepackage{dashrule}
\usepackage{multirow}
\usepackage{multicol}
\usepackage{mathtools}
\usepackage[algoruled,resetcount,linesnumbered]{algorithm2e}
\usepackage{lipsum}
\newtheorem{lemma}{Lemma}

\newtheorem{remark}{Remark}

\usepackage{caption}
\usepackage{comment}
\newcolumntype{P}[1]{>{\centering\arraybackslash}p{#1}}
\usepackage{lipsum}

\newcommand\blfootnote[1]{%
  \begingroup
  \renewcommand\thefootnote{}\footnote{#1}%
  \addtocounter{footnote}{-1}%
  \endgroup
}

\graphicspath{{Figures/}}
\title{Channel Estimation in mmWave Hybrid MIMO System via 
Off-Grid Dirichlet Kernels}
\author{Chethan Kumar Anjinappa, You Zhou, Yavuz Yapici, Dror Baron, and Ismail Guvenc\\
\IEEEauthorblockA{Department of Electrical and Computer Engineering, 
North Carolina State University, Raleigh, NC
}
Email: {\tt \{canjina, yzhou26, yyapici, dzbaron, iguvenc\}@ncsu.edu}}

\begin{document}

\maketitle

\begin{abstract}
In this paper, we tackle channel estimation in millimeter-wave hybrid multiple-input multiple-output systems by considering off-grid effects. 
In particular, we assume that spatial parameters can take any value in the angular domain, and need not fall on predefined discretized angles. Instead of increasing the number of discretized points to combat off-grid effects, we use implicit \emph{Dirichlet kernel} structure in the Fourier domain, which conventional compressed sensing methods do not use. We propose greedy low-complexity algorithms based on orthogonal matching pursuit (OMP);
our core idea is to traverse the Dirichlet kernel peak using estimates of the discrete Fourier transform. We demonstrate the efficacy of our proposed algorithms compared to standard OMP reconstruction. Numerical results show that our proposed algorithms obtain smaller reconstruction errors when off-grid effects are accounted for.
\blfootnote{This research was supported in part by NSF under the grant numbers ACI-1541108 and ECCS-1611112.}
\end{abstract}


\section{Introduction and Related Work}
One of the most promising features of next-generation wireless systems is to use high-frequency high-bandwidth signals in millimeter-wave (mmWave) frequency bands. These mmWave bands combined with multiple-input multiple-output (MIMO) technology have great potential in delivering higher data rates, higher spectral efficiency, and lower latency, exceeding the performance of traditional cellular systems operating at sub-6 GHz bands. 
Conventional mmWave MIMO architectures use a large number of antennas, which results in high cost and power consumption, 
making it difficult to assign a radio frequency (RF) chain per antenna. To curtail these issues, a \textit{hybrid analog/digital beamforming (HADB)} architecture is adapted at mmWave bands \cite{mendez2016hybrid,heath2016overview}. 

The HADB architecture complicates the channel estimation process, because only the low dimensional signals pre-combined by the analog combiner are available at baseband, which severely degrades the channel estimation process. The accuracy with which the channel is estimated plays a critical role in physical layer performance as it directly affects receiver design, e.g., channel equalization \cite{SSR_Equalization} and radio resource management \cite{RRM}. 
To overcome these challenges, channel estimation algorithms based on compressed sensing (CS) \cite{Uniform_Sampling_Cos_Domain,heath2016overview} 
have been proposed. These CS-based methods are based on \textit{virtual channel models} \cite{sayeed2002deconstructing}, which provide a virtual angular representation of MIMO channels. 

The virtual channel model describes the channel with respect to (w.r.t.) fixed basis functions corresponding to spatial angles within a finite discrete dictionary. In other words, the continuous parameter space of spatial angular features is discretized into a finite set of pre-defined spatial angles, which emphasizes the sparse representation of the MIMO channels. The estimation accuracy of CS methods based on this discretization is limited by the number of points in the dictionary. Although this discretization procedure yields state-of-art performance, it has several intrinsic disadvantages \cite{tang2013compressed_off_grid}, including the \textit{off-grid effect}.

A natural yet inefficient approach to reduce off-grid effects is to increase the number of discretized points, 
corresponding to increased grid resolution. This approach not only increases the mutual coherence of the dictionary matrix, 
leading to loss of the restricted isometric property, but also increases the problem dimension, which requires more computation \cite{PPOMP}. 
An alternative is to tackle off-grid effects upfront without increasing the grid size. For example, in the context of channel estimation, Gurbuz et al. \cite{PPOMP} provide a controlled perturbation mechanism for spatial angular parameters based on orthogonal matching pursuit (OMP) \cite{OMP}. 
Other related works involve an improved off-grid sparse Bayesian algorithm \cite{tang2019off}, and  
a grid-less CS technique developed via atomic norm minimization 
\cite{2017_Gridless_Atomic}. 
Although these methods all tackle off-grid issues, 
they are computationally prohibitive, which motivates us to develop and analyze robust low-complexity channel estimation algorithms that 
account for off-grid effects. 

Interestingly, standard CS methods based on sparsity fail to leverage {\em Dirichlet structure} in the Fourier domain. We exploit this structure 
to improve the channel estimation process. In particular, we propose low-complexity algorithms based on OMP \cite{OMP}, owing to its computational tractability. Our numerical results show that while accounting for off-grid effects, our proposed algorithms obtain smaller channel reconstruction errors compared to standard OMP algorithms.

{\bf Notation}: Vectors and matrices are represented by lower-case  
and capital boldface letters, respectively (e.g.: \textbf{a} and \textbf{A}). The transpose, conjugate, conjugate transpose, and pseudo-inverse of a matrix \textbf{A} are denoted by  $\mathbf{A}^\text{T}$, $\mathbf{A}^\text{H}$, $\mathbf{A}^{*}$, and $\mathbf{A}^{\dagger}$, respectively. For a non-negative integer $K$, we denote the set $\{1,2,\ldots, K\}$ by $[K]$; $\otimes$ denotes the Kronecker product; vec($\mathbf{A}$) denotes the vectorized version of the matrix $\mathbf{A}$; $\mathbb{R}$ is the real part of a complex number; and $\min(a,b)$ is the minimum of the scalars $a$ and $b$.

\section{System Model and Channel Model}
\subsection{System Model}
Consider a mmWave MIMO network comprised of a base station (BS) communicating with generic user equipment (UE), both equipped with a uniform linear array (ULA). We assume the BS is equipped with $M$ antennas, $M_\text{RF}$ RF chains, and $M_\text{DS}$ data streams. Similarly, the UE is assumed to be equipped with $N$ antennas, $N_\text{RF}$ RF chains, and $N_\text{DS}$ data streams. Typically, it is assumed that $M_\text{DS} \leq M_\text{RF}\leq M$ and $N_\text{DS} \leq N_\text{RF}\leq N$. With the HADB MIMO processing structure \cite{mendez2016hybrid}, the received signal at the UE is expressed,
\begin{equation}\label{y_Equ}
\mathbf{Y} = \mathbf{W}^\text{H} \mathbf{H F s} + \mathbf{N}_{W},
\end{equation}
where $\mathbf{Y}$ $\in \mathcal{C}^{N_\text{t}\times M_\text{t}}$ is the received measurement matrix at the UE, and 
it is assumed that the UE uses $N_\text{t}$ combiners for each $M_\text{t}$ beamforming vector used by the BS. 
The received measurements at the UE for each beamforming vector are arranged in columns. 
The matrices $\mathbf{H}$ $\in \mathcal{C}^{N \times M}$,  $\mathbf{W}$ $\in \mathcal{C}^{N \times N_\text{t}}$, and $\mathbf{F}$ $\in \mathcal{C}^{M \times M_\text{t}}$ represent the channel matrix from the BS to UE, 
the combined effect of the RF/baseband combiner,
and precoder matrices, respectively. The noise matrix at the UE after the combiner operation is $\mathbf{N}_{W} = \mathbf{W}^\text{H} [\mathbf{n}_1, \ldots, \mathbf{n}_{M_\text{t}}]$, where $\mathbf{n}_\text{i} \in \mathcal{C}^{N \times 1}; \forall \text{i} \in [M_\text{t}]$, follows a circularly symmetric  independent and identically distributed Gaussian distribution, $\mathcal{CN}(0,\sigma_{n}^2)$ with noise variance~$\sigma_{n}^2$. Further, $\mathbf{s}$ is assumed to be known at the BS and omitted hereafter. 

\subsection{Channel Model}
Based on \cite{ChannelModel_Rappaport}, the mmWave channel model is given by:
\begin{equation}\label{H_Equation}
\mathbf{H} = \sum_{l=1}^L \alpha_l \mathbf{a}_\text{UE}(\theta_l) \mathbf{a}_\text{BS}^\text{H}(\phi_l),
\end{equation}
where $\alpha_l$ is the complex gain associated with $l^\text{th}$ multi-path component (MPC) between the BS and UE. The number of MPCs is $L$ with $L \,{\ll}\, {M;N}$ may itself be time-varying due to the mobility of the UE and the surrounding scatterers \cite{Angular_Temporal_Correlation_Chethan_2018}. The terms $\mathbf{a}_\text{BS}(\phi)$ and $\mathbf{a}_\text{UE}(\theta)$ are the normalized array response to an MPC coming from the angles $\phi$ and $\theta$ w.r.t. the BS and UE ULA, respectively. The tuple $(\phi,\theta)\in [-\pi/2,\pi/2]$ is the physical azimuth angle-of-departure (AoD) and angle-of-arrival (AoA), respectively. The normalized ULA responses at the BS and UE are expressed as 
\begin{eqnarray}\label{a_BSa_UE}
\begin{aligned}
&[\mathbf{a}_\text{BS} (\phi)]_m= \frac{1}{\sqrt{M}} e^{j \frac{2 \pi }{\lambda}d_\text{BS}(m-1) \sin(\phi)}, \quad m \in [M],\\
&[\mathbf{a}_\text{UE}(\theta)]_n = \frac{1}{\sqrt{N}} e^{j \frac{2 \pi}{\lambda}d_\text{UE}(n-1) \sin(\theta)}, \quad n \in [N],\vspace{-.25cm}
\end{aligned}
\end{eqnarray}
where $d_\text{BS}$ and $d_\text{UE}$ are the inter-element spacings in the BS and UE ULA, respectively. We assume $d_\text{BS}= d_\text{UE} =\frac{\lambda}{2}$ where $\lambda$ is the carrier wavelength defined by $\frac{c}{f_0}$ with $c$ and $f_0$ being the speed of light and the carrier frequency, respectively. 
\begin{figure*}[!t]
   \centerline{ \includegraphics[scale=0.4]{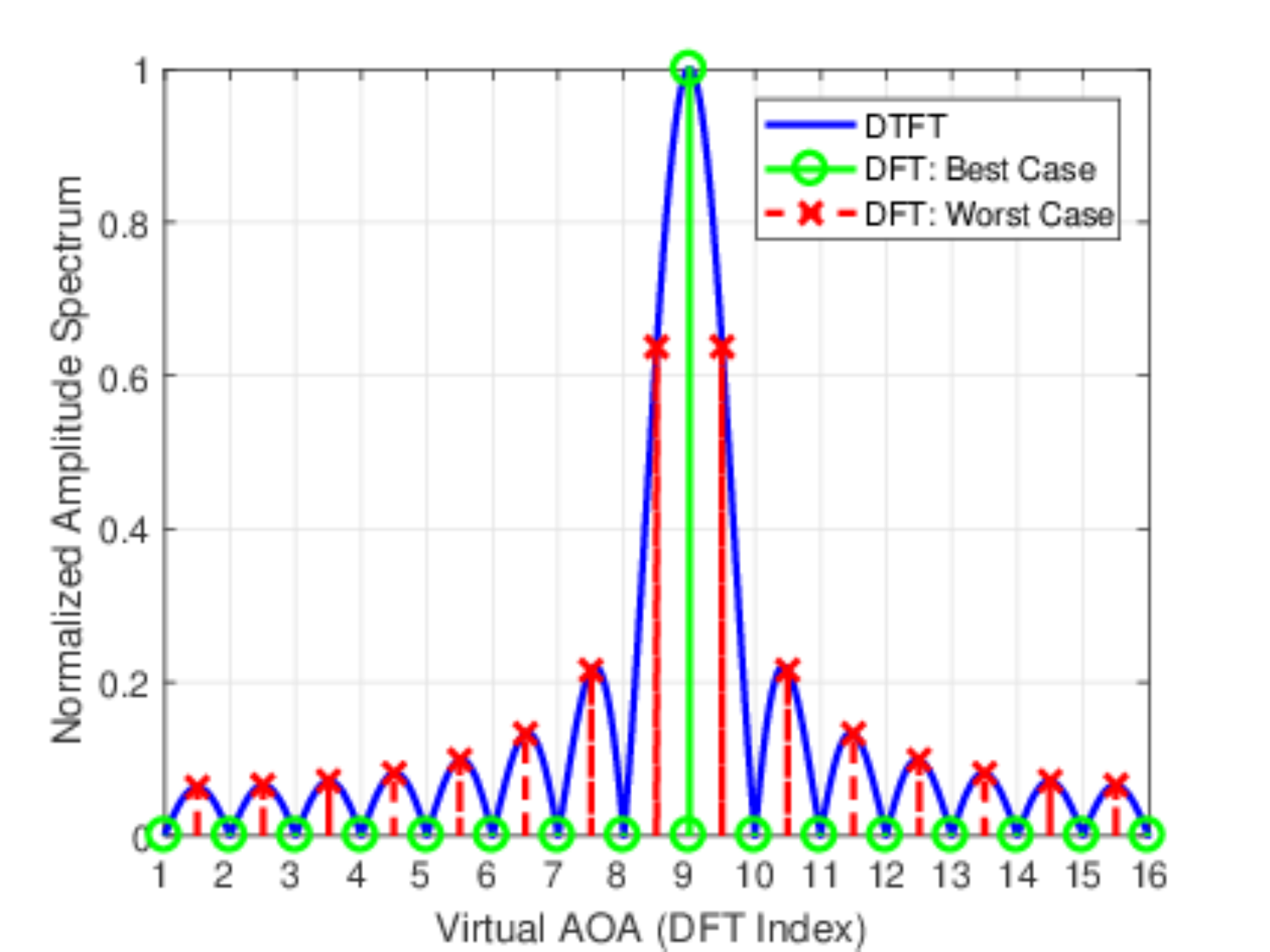} \includegraphics[scale=0.4]{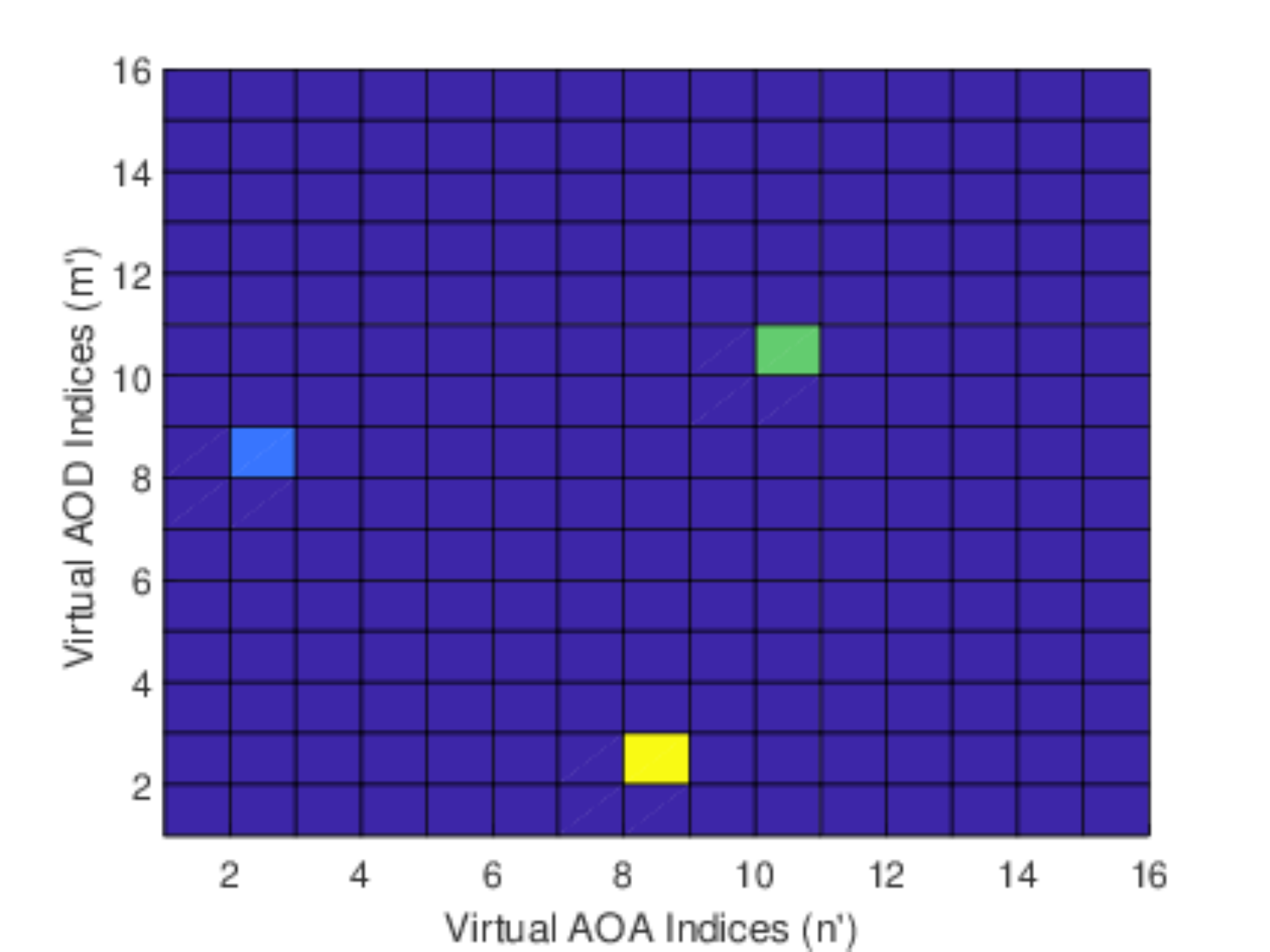}\includegraphics[scale=0.4]{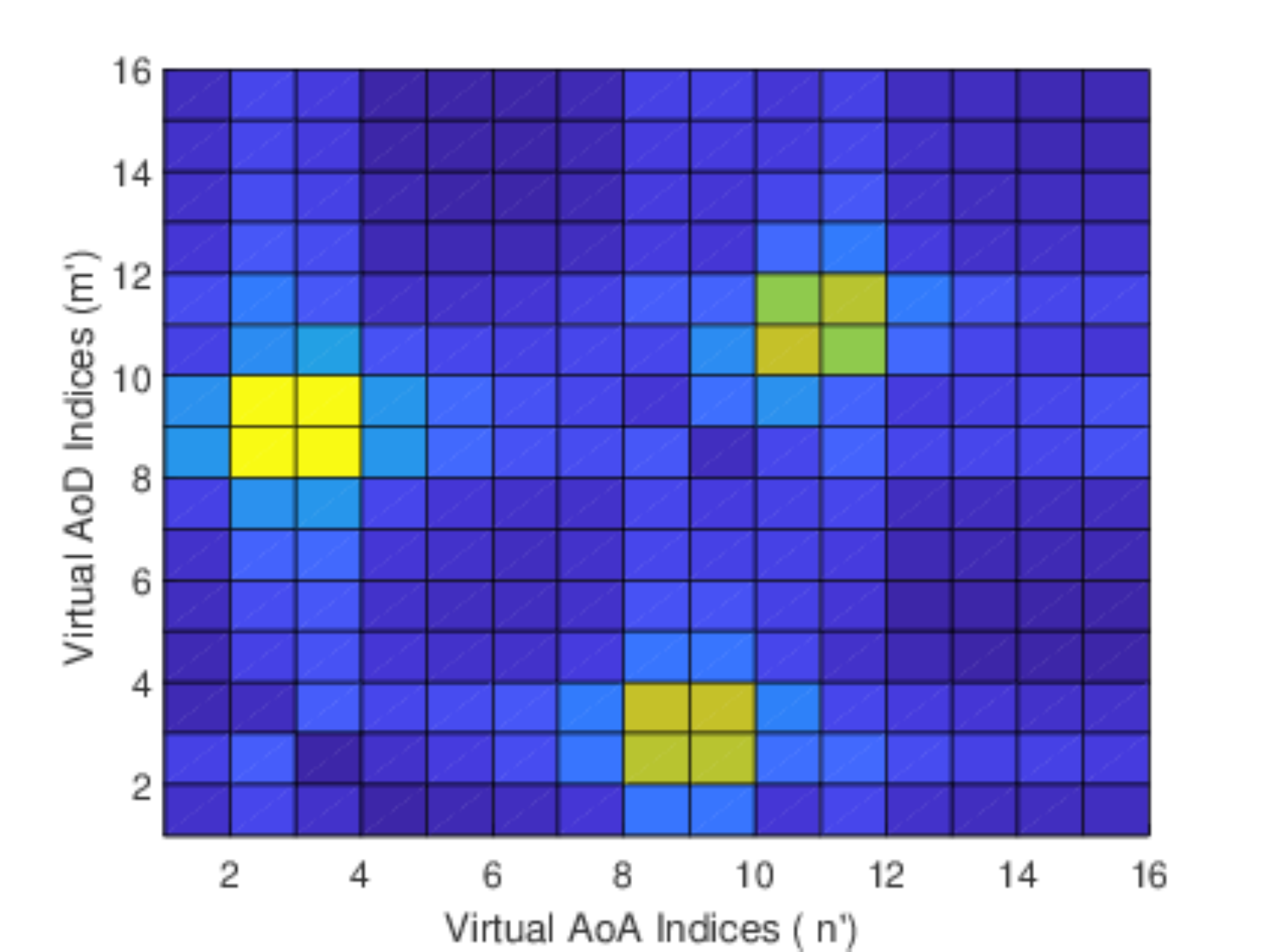}}    \vspace{-1mm}
        \caption{\textbf{(Left)} Normalized DTFT and DFT amplitude spectrum of the virtual beampscae matrix of a single MPC in the spatial AoA domain with $M = N = 16$. \textbf{(Middle)} On-Grid and \textbf{(Right)} worst off-grid effect visualization in the 2D-virtual domain with three unit strength MPCs for $M = N = 16$. The ideal on-grid case results in exact sparse representation in the virtual domain as the DFT and the Dirichlet (DTFT) peak coincide, whereas, in the worst off-grid condition the DFT and the Dirichlet peaks do not coincide resulting in a significant increase in the number of non-zero elements.}
    \label{fig:2D_Dirichlet}
    \vspace{-4mm}
\end{figure*}

\subsection{Sparse Beamspace (Virtual) Representation}
We use the \textit{virtual channel} model representation of $\mathbf{H}$, which relates the beamspace and antenna space by the spatial Fourier transform. The virtual channel model describes the channel w.r.t. fixed basis functions corresponding to spatial angles from the finite discrete dictionary. In particular, we follow the framework in Lee et al. \cite{Uniform_Sampling_Cos_Domain} and discretize the tuple $(\hat{\phi},\hat{\theta})$ such that the $(\sin(\hat{\phi}),\sin(\hat{\theta}))$ appearing in the array responses (\ref{a_BSa_UE}) are uniformly distributed in [-1,1). Specifically, the quantized grids should satisfy:
\begin{eqnarray}\label{Eq:Discretization}
\begin{aligned}
&\mathbf{\Phi} = \{\hat{\phi}: \frac{1+\sin(\hat{\phi})}{2}=\frac{m-1}{M}; m \in [G_\text{BS,D}] \},\\
&\mathbf{ \Theta} = \{\hat{\theta}: \frac{1+\sin(\hat{\theta})}{2} = \frac{n-1}{N}; n \in [G_\text{UE,D}]\},
\end{aligned}
\end{eqnarray}
where $G_\text{UE,D}$ and $G_\text{BS,D}$ are the grid size for the spatial AoA-AoD, respectively. The array response corresponding to the discretized spatial angles are grouped to form the matrices ${\mathbf{A}_\text{BS,D}}=\{\mathbf{a}_\text{BS}(\hat{\phi}); \hat{\phi} \in \mathbf{\Phi}\}$ and ${\mathbf{A}_\text{UE,D}} =\{\mathbf{a}_\text{UE} (\hat{\theta});\hat{\theta} \in \mathbf{\Theta}\}$, which are transmitting and receiving beamforming matrices at BS and UE, respectively. Generally, $G_\text{BS,D} \geq M$ and $G_\text{UE,D} \geq N$. However, throughout this work 
we assume $G_\text{BS,D} = M$ and $G_\text{UE,D}=N$, resulting in 
$\mathbf{A}_\text{BS,D} \in \mathcal{C}^{M \times M}$ and $\mathbf{A}_\text{UE,D}\in \mathcal{C}^{N \times N}$ being unitary Inverse-discrete Fourier transform (DFT) matrices,
which are represented as 
\begin{eqnarray}
\begin{aligned}
&[\mathbf{A}_\text{BS,D}]_{m,m'} = \frac{1}{\sqrt{M}} e^{j 2\pi (m-1) (\frac{m'-1}{M} -\frac{1}{2})};\quad m, m' \in [M],\\
&[{\mathbf{A}}_\text{UE,D}]_{n,n'} = \frac{1}{\sqrt{N}} e^{j 2\pi (n-1) (\frac{n'-1}{N} -\frac{1}{2})};\quad n, n' \in [N].
\end{aligned}
\end{eqnarray}
The exact beamspace representation can be expressed as $\mathbf{A}_\text{UE,D} \mathbf{H}_\text{V} \mathbf{A}_\text{BS,D}^\text{H}$,
where $\mathbf{H}_\text{V} \in \mathcal{C}^{N \times M}$ is the beamspace sparse matrix defined as follows:
\begin{eqnarray}\label{Eq:Hv_Eq}
\begin{aligned}
\mathbf{H}_\text{V}  &= \mathbf{A}_\text{UE,D}^\text{H} \mathbf{H} \mathbf{A}_\text{BS,D} = \sum_{l=1}^L \alpha_l \mathbf{\hat{a}}_\text{UE}(\theta_l) \mathbf{\hat{a}}_\text{BS}^\text{H}(\phi_l),
\end{aligned}
\end{eqnarray}
and $\mathbf{\hat{{a}}}_\text{UE}(\theta)$ = $\mathbf{A}_\text{UE,D}^\text{H} \mathbf{a}_\text{UE}(\theta)$ and 
$\mathbf{\hat{{a}}}_\text{BS}(\phi)$ = $\mathbf{A}_\text{BS,D}^\text{H}\mathbf{a}_\text{BS}(\phi)$ are the normalized UE and BS array responses 
w.r.t. the DFT basis, respectively. The UE array response w.r.t. the DFT basis can be compactly represented as \cite{tang2013compressed_off_grid,song2018scalable},
\begin{eqnarray}\label{Eq:a_UE}
\begin{aligned}
    \big[\mathbf{\hat{a}}_\text{UE}(\theta_l)\big]_{n'} 
    &=\frac{1}{{N}} \sum_{i=0}^{N-1} e^{j2\pi i (\frac{n' - 1}{N} - \frac{1}{2})} e^{j \pi i \sin{\theta_l}}\\
   & = \frac{1}{{N}} \frac{\sin(\pi \vartheta_{n',l} N)}{\sin{(\pi \vartheta_{n',l})}} e^{-j \pi \vartheta_{n',l} (N-1)},
\end{aligned}
\end{eqnarray}
where $\vartheta_{n',l} = \frac{n'-1}{N} - \frac{1}{2}\sin{(\theta_l)} -\frac{1}{2}; n' \in [N]$. The proof of the equivalent representation is straightforward and omitted for brevity. Similarly, $\mathbf{\hat{a}}_\text{BS}(\phi_l)$ is formulated as 
$ [\mathbf{\hat{a}}_\text{BS}(\phi_l)]_{m'} = \frac{1}{{M}} \frac{\sin(\pi \varphi_{m',l} M)}{\sin{(\pi \varphi_{m',l})}} e^{-j \pi \varphi_{m',l} (M-1)}$, where $\varphi_{m',l} = \frac{m'-1}{M} - \frac{1}{2}\sin{(\phi_l)} -\frac{1}{2}; m' \in [M]$. 
Substituting $\mathbf{\hat{a}}_\text{UE}$ and $\mathbf{\hat{a}}_\text{BS}$ into (\ref{Eq:Hv_Eq}), the $(m',n')$ entry of the beamspace matrix $\mathbf{H}_\text{V}$ becomes
\begin{eqnarray}\label{Eq:Hv1}
\begin{aligned}
\big[\mathbf{H}_\text{V} \big]_{m',n'} =\sum_{l=1}^L \alpha_l\mathcal{D}(\varphi_{m',l},\vartheta_{n',l}) \frac{e^{-j \pi\vartheta_{n',l} (N-1)}}{e^{-j \pi \varphi_{m',l} (M-1)}},
\end{aligned}
\end{eqnarray}
where $\mathcal{D}(\varphi_{m',l},\vartheta_{n',l}) = \frac{1}{{MN}}\frac{\sin(\pi \varphi_{m',l} M)}{\sin{(\pi \varphi_{m',l})}}\frac{\sin(\pi \vartheta_{m',l} N)}{\sin{(\pi \vartheta_{m',l})}}$
is the 
\textit{{Dirichlet kernel}} (note that $\mathcal{D}(\varphi_{m',l},\vartheta_{n',l}) = 1$ when $\varphi_{m',l} = \vartheta_{n',l}= 0$). 
Since the DFT is discrete in nature, the beamspace domain $\mathbf{H}_\text{V}$ in (\ref{Eq:Hv1}) is evaluated only at integer points
($m',n'$) with $m'\in [M]$, $n'\in [N]$. However, the Dirichlet kernel peak need not occur at any of these integer points ($m'$, $n'$) and
can take continuous values, i.e.,  $m^{\star} \in [1,M], n^{\star} \in [1,N]$. Therefore, the Dirichlet kernel need not peak at one of the 
pre-defined spatial angles, and maxima of the DFT may not correspond to maxima of the Dirichlet kernel 
or maxima of the discrete-time Fourier transform (DTFT). 

We now sidestep away from the DFT representation to discuss the DTFT concept, which will be pivotal in understanding 
our algorithms
in Section \ref{Sec:3}. The continuum of (\ref{Eq:Hv1}) evaluated at $m^{\star} \in [1,M]$ and $n^{\star} \in [1,N]$ 
is represented as 
\begin{eqnarray}\label{Eq:Hv2}
\begin{aligned}
[\mathbf{H}_\text{V}]_{m^\star,n^\star} =\sum_{l=1}^L \alpha_l\mathcal{D}(\varphi_{m^\star,l},\vartheta_{n^\star,l})  \frac{e^{-j \pi\vartheta_{n^\star,l} (N-1)}}{e^{-j \pi \varphi_{m^\star,l} (M-1)}}.
\end{aligned}
\end{eqnarray}
This form (\ref{Eq:Hv2}) is the DTFT counterpart of (\ref{Eq:Hv1}), where $(m^\star,n^\star)$ can take continuous values. 
That is, $1\leq m^\star \leq M$ and $1\leq n^\star \leq N$. Intuitively, each MPC in the physical domain results in a continuous Dirichlet 
kernel in the continuum of the beamspace domain, as shown in Fig. \ref{fig:2D_Dirichlet}. 

\subsection{Sparse Recovery Problem}
Aided by the sparse virtual representation and vector identity property, vec($\bf{ABC}$) = $(\mathbf{C}^T \otimes \mathbf{A}) \text{vec}(\mathbf{B})$, MIMO channel estimation (\ref{y_Equ}) is posed as sparse recovery \cite{mendez2016hybrid}
and rewritten:
\begin{equation}\label{SSR_Setup}
    \mathbf{y} = \mathbf{A} {\text{vec}([\mathbf{H}_\text{V}]_{m',n'})} + \mathbf{n}_{W},
\end{equation}
where $\mathbf{A}= \mathbf{\Phi \Psi} \in \mathcal{C}^{M_\text{t} N_\text{t} \times M N}$ is the overall sensing matrix, 
$\mathbf{\Phi} = (\mathbf{F}^\text{T} \otimes \mathbf{W}^\text{H})$ represents the combined effect of the precoder and combiner,
$\bf{\Psi} = (\mathbf{A}_\text{BS,D}^* \otimes \mathbf{A}_\text{UE,D})$ is the dictionary matrix, and 
$\bf{A_\text{BS,D}}$ and $\mathbf{A}_\text{UE,D}$ are matrices for the DFT basis. 
Finally, $\mathbf{y} \in \mathcal{C}^{M_\text{t}N_\text{t} \times 1}$ is the vectorized form of $\mathbf{Y}$. 

Conventional CS techniques assume that the signal is exactly sparse, which is true only when the physical AoA-AoD tuples are aligned 
with discretized spatial angles; this is the ideal on-grid case. However, the physical AoA-AoD ($\theta_l, \phi_l$) take continuous values, 
which may not be aligned with any discretized spatial angles, resulting in basis mismatch (off-grid) effects. 
These effects violate the sparsity assumption, resulting in performance degradation of CS-based techniques \cite{tang2013compressed_off_grid}. 
Below, we analyze the impact of basis mismatch on the sparsity level, which is central to any CS methods. 

\subsection{Effect of Off-Grid/Basis Mismatch}\label{Basis_Mismatch}
\subsubsection*{Best Case (on-grid)} The physical AoA-AoD ($\theta_l,\phi_l$) tuple falls exactly on any 
of the pre-defined spatial angles, and the DFT and DTFT peaks coincide, resulting in exact sparse representation. 
This phenomenon is illustrated by a 1-dimensional (1D) normalized Dirichlet kernel (Fig. \ref{fig:2D_Dirichlet}, left). 

\subsubsection*{Worst Case (off-grid)} The physical AoA-AoD ($\theta_l,\phi_l$) tuple results in 
a virtual AoA-AoD ($\varphi_l,\vartheta_l$), which is exactly in between any adjacent pre-defined virtual angles. That is, the resulting virtual AoA-AoD tuple is at distance $(\frac{1}{2M},\frac{1}{2N})$ away from some predefined virtual angle. In this case, the DTFT and DFT peaks do not coincide,
which  affects not only the two adjacent cells but the entire grid, with amplitude decaying at rate $1/M$ and $1/N$ in the virtual AoA-AOD, respectively, violating the sparsity assumption. Examples of on-grid and worst off-grid effects in a MIMO system appear in Fig. \ref{fig:2D_Dirichlet}
(middle and right panels, respectively).
 
\begin{figure}[!t]\vspace{-.2cm}
    \centerline{\includegraphics[scale=0.65]{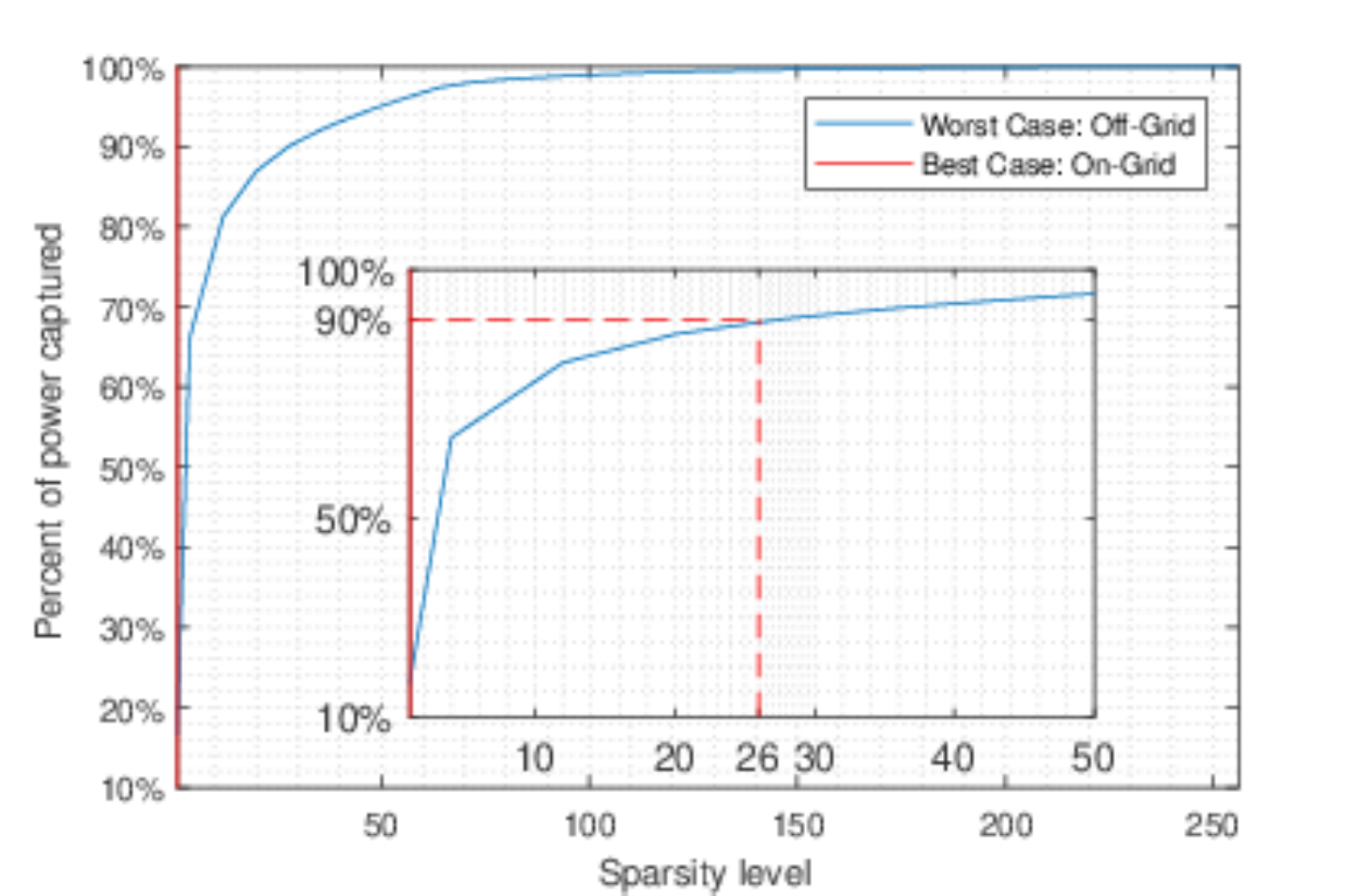}}
\vspace{-1mm}    \caption{Best and Worst Case Scenarios: Power captured by $K$ dominant elements in the virtual domain with a single MPC for $M = N = 16$.}
    \label{fig:Best_Worst}
\end{figure}

To provide insights about the worst case of the off-grid problem, we provide a Lemma for MIMO in the presence of a single unit strength MPC,
which resembles Lemma 2 in Gao et al. \cite{gao2017fast}. Our Lemma provides insights about the number of non-zero indices 
(which we define as the \emph{sparsity level}) that need to be recovered by the CS methods to capture $\eta$ percent of the power of an MPC in the virtual domain. We concentrate on the worst-case and upper bound the general case, which can be extended to multiple MPCs as well.
\begin{lemma}
 Let $K$ represent the sparsity level in the virtual AoA-AoD domain. Without loss of generality, we assume $K$ to be a multiple of 4. 
 The power captured by the $K$ strongest elements in the DFT domain is given by $\eta = \frac{P_K}{P_T}$, where
 \begin{eqnarray}
 \begin{aligned}
    P_K &= \frac{4}{{M N}} \sum_{i=1}^{K/4} \sum_{j=1}^{K/4}  \left| \frac{\sin(M\pi\frac{(2i-1)}{2})}{\sin(\pi\frac{(2i-1)}{2})} \frac{\sin(N\pi\frac{(2j-1)}{2})}{\sin(\pi\frac{(2j-1)}{2})}\right|^2,\\
    P_T & =\frac{4}{{M N}} \sum_{i=1}^{M/4} \sum_{j=1}^{N/4}  \left| \frac{\sin(M\pi\frac{(2i-1)}{2})}{\sin(\pi\frac{(2i-1)}{2})} \frac{\sin(N\pi\frac{(2j-1)}{2})}{\sin(\pi\frac{(2j-1)}{2})}\right|^2.
    \end{aligned}
\end{eqnarray}
\end{lemma}
The ratio $\eta$ is the power captured by $K$ dominant $K$ DFT elements. 
For the worst and best case scenarios, Fig. \ref{fig:Best_Worst} shows 
the number of non-zeros that must be recovered by CS algorithms 
in order to capture $\eta$ power for a single MPC with $M = N = 16$. In other scenarios, the sparsity level is between the 
blue and red lines. As evident from the figure, in the worst case scenario CS algorithms must recover significantly more non-zeros,
which is inefficient. A better way to capture power in the virtual domain requires that we reach the peak of the DTFT spectrum instead of 
operating on DFT peaks. In other words, we can capture all the power (100\%) 
of each MPC by identifying the peak location of the Dirichlet kernel. 

In light of these observations, our main objective is to find the maxima of the Dirichlet kernel peaks (strength/location) 
instead of recovering all the non-zero elements in the virtual DFT domain. Mathematically, the peak strength and location can be obtained by solving the following optimization,
\begin{eqnarray}\label{Eq:Joint_Opt}
\begin{aligned}
&\hat{\alpha_l}^\star,\hat{m}_l^\star,\hat{n}_l^\star =  \underset{\alpha_l^\star,{m}_l^\star,{n}_l^{\star}}{\text{min}}  ||\mathbf{y} - \mathbf{A} \text{vec}([\mathbf{H_\text{V}}]_{m',n'}) ||_2^2
\end{aligned}
\end{eqnarray}
\vspace{-.55cm}
\begin{eqnarray*}\label{Eq:Joint_Opt1}
\begin{aligned}
&\textit{s.t.} \quad  1\leq m, m_l^\star \leq M, 1 \leq  n, n_l^\star  \leq N, m'\in [M], n' \in [N],\\
& \scriptsize{\mathbf{H}_\text{V} \hspace{-.1cm}=\hspace{-.1cm} \sum_{l=1}^L  \frac{\alpha_l^\star}{{MN}}\hspace{-.1cm}\frac{\sin(\pi (m - m_l^{\star}) )\sin(\pi (n - n_l^{\star}))e^{-j\pi(m-m_l^\star)}}{\sin{(\frac{\pi}{M} (m - m_l^{\star}))}\sin{(\frac{\pi}{N}( n - n_l^{\star}))}e^{-j\pi(n-n_l^\star)}}}.
\end{aligned}
\end{eqnarray*}
Note that $\mathbf{H}_\text{V}$ in our objective function is still evaluated at integer points (${m',n'}$). This optimization procedure 
finds the strength and location of the Dirichlet kernel peaks while minimizing the $\ell_2$ norm of the residual between the estimated 
parameters and measurement vector. The constraints mandate that the search is not just over a finite set of angles defined in (\ref{Eq:Discretization}) 
but over the entire space. Intuitively, finding maxima locations of the Dirichlet kernel (DTFT spectrum) corresponds to
estimating the AoA-AoD ($\theta_l,\phi_l$). Similarly, finding the peak strength of the Dirichlet kernel corresponds to estimating the strength of the MPC $\alpha_l$. 

The above joint optimization problem (\ref{Eq:Joint_Opt})
is non-convex and in general challenging. The non-convexity mainly arises from the Dirichlet structure; Fig.\ref{fig:2D_Dirichlet} exudes it. For brevity, we omit the proof of non-convexity. Below, we propose simpler and efficient schemes to solve (\ref{Eq:Joint_Opt}) using coarse estimates of DFT points obtained from the greedy OMP algorithm.

\begin{figure}[!t]
    \centerline{
    \includegraphics[scale=0.6]{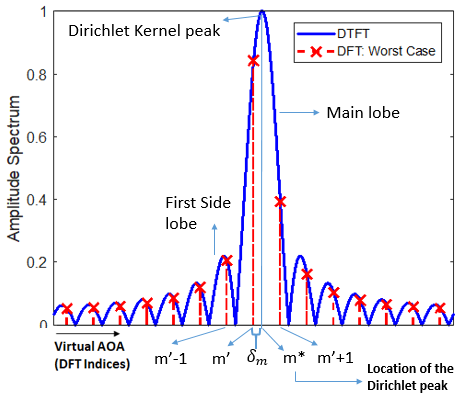}}\vspace{-3mm}
    \caption{DFT and DTFT amplitude spectrum for a single MPC in the virtual AoA domain.}
    \label{fig:Toy_Dirichlet}
\end{figure}

\section{Exploiting Dirichlet Kernel Structure}\label{Sec:3}
This section begins by investigating the Dirichlet kernel for a single MPC. After addressing a single MPC, 
we propose an algorithm that accommodates MIMO and  multiple MPCs.


To keep our presentation simple, suppose that a single MPC falls off the grid, resulting in a Dirichlet kernel in the virtual AoA domain
(Fig. \ref{fig:Toy_Dirichlet}). Denote the DFT and Dirichlet kernel peaks by $m'$ and $m^{\star}$, respectively. Due to properties 
of the DFT and DTFT, the Dirichlet kernel peak lies in 
the range $[m'-1, m'+1]$. We compute least square (LS) estimates at locations $\{m'-1,m',m'+ 1\}$ (integer indices adjacent to $m'$)
under the constraint of Dirichlet structure, resulting in estimates $[\mathbf{H}_\text{V}]_{m'-1}$, $[\mathbf{H}_\text{V}]_{m'}$, and $[\mathbf{H}_\text{V}]_{m'+1}$. The goal is to traverse to the Dirichlet kernel peak and reconstruct it using the LS estimates. 
This can be achieved in at least two ways.

\subsection{Dirichlet OMP-Main Lobe (DOMP-MLb)} 
Without loss of generality, suppose that the Dirichlet peak is in the range $[m', m'+1]$. In this case, $|[\mathbf{H}_\text{V}]_{m'+1}| > |[\mathbf{H}_\text{V}]_{m'-1}|$, implying that the main-lobe (MLb) is in the range $[m', m'+1]$. Thus, we can ignore the side lobe (SLb) 
estimate. Based on the MLb estimates, we can find the location $m^\star  = m' + \delta_m$, where  
$\delta_m = \frac{1}{2} \min\left( \frac{[\mathbf{H}_\text{V}]_{m'}}{[\mathbf{H}_\text{V}]_{m'+1}}, \frac{[\mathbf{H}_\text{V}]_{m'+1}}{[\mathbf{H}_\text{V}]_{m'}}\right)$ 
is the deviation from the DFT index, which is obtained by exploiting the uni-modal symmetric, concave property of the main lobe 
(Fig.~\ref{fig:Toy_Dirichlet}). 
\subsection{DOMP-Main and Side Lobe (DOMP-MSLb)} 
Instead of considering just the MLb estimate, the algorithm becomes more robust by
not discarding the SLb estimates, but using them to estimate the Dirichlet peak.
Based on the main and side lobe (MSLb) estimates, we can find the location $m^\star  = m' + \delta_m$, where $\delta_m = \scriptsize{\frac{\tan(\frac{\pi}{M})}{\frac{\pi}{M}} \mathbb{R}\left( \frac{[\mathbf{H_\text{V}}]_{m'-1}  - [\mathbf{H}_\text{V}]_{m'+1}}{2[\mathbf{H}_\text{V}]_{m'} - [\mathbf{H}_\text{V}]_{m'-1} - [\mathbf{H}_\text{V}]_{m'+1}} \right)}$; details in \cite{candan2011method}.

Extending the previous ideas to a 2D MIMO problem is straightforward. For the 2D problem, suppose that the DFT  peaks occur at ($m',n'$), where $m'$ and $n'$ refer to indices in the virtual AoA/AoD domains, respectively. Similarly, the Dirichlet kernel peak occurs at ($m^{\star},n^{\star}$). For each MPC in the MIMO case, we need 5 estimates: the DFT peak index ($m',n'$) and LS estimate ($[\mathbf{H}_\text{V}]_{m',n'}$) and 4 DFT indices and LS estimates around the DFT peak (denoted by $\kappa$ and $[\mathbf{H}_\text{V}]_{\kappa}$ in Algorithm \ref{Algorithm:Dirichlet_Alg1}, respectively). With this information, one can reconstruct the 2D Dirichlet and repeat the procedure recursively for each MPC (denoted by $l$). Further details of our proposed reconstruction method are summarized in Algorithm~\ref{Algorithm:Dirichlet_Alg1}; some remarks are in order.
\begin{figure*}[t!]
\end{figure*}
\begin{algorithm}[t!]
\caption{Channel Estimation: DOMP-MLb/MSLb} \label{Algorithm:Dirichlet_Alg1}
\KwIn{$\mathbf{y}$, $\mathbf{A}$, $\epsilon$ $\quad$\\ Initialization: $\mathbf{\mathcal{S}} = \{ \}$, $\mathbf{y}_\text{res} = \mathbf{y}$, $e  = || \mathbf{y}_\text{res}||_2$, $l$ = 1.}
\While{$e < \epsilon$}{
{
$j^{\star} = \arg \underset{j}{\max}| \mathbf{A}(:,j)^\text{T} \mathbf{y}_\text{res}|$;\quad $\mathcal{S} = \{j^\star\pm M, j^\star\pm N\}$\\
$m'$ = floor(${j^{\star}}/{M}$); \quad  $n'$ = mod(${j^{\star}},{N})\quad$\\
$\kappa \in \{(m',n'),(m'\pm 1,n'),(m',n'\pm 1)\}$\\
$[\mathbf{H}_\text{V}]_{\kappa} = (\mathbf{A}(:,\mathcal{S}))^{\dagger}$ $\mathbf{y}_\text{res}$
\\
\eIf{\text{\textit{DOMP-MLb Update:}}}{
 \eIf{$|[\mathbf{H}_\text{V}]_{m'+1,n}| > |[\mathbf{H}_\text{V}]_{m'-1,n}|$}{
$m^\star_l = m' + \frac{1}{2} \min\left( \frac{[\mathbf{H}_\text{V}]_{m',n'}}{[\mathbf{H_\text{V}}]_{m'+1,n'}}, \frac{[\mathbf{H}_\text{V}]_{m'+1,n'}}{[\mathbf{H}_\text{V}]_{m',n'}}\right)$
        }{
$m_l^\star = m' - \frac{1}{2} \min\left( \frac{[\mathbf{H}_\text{V}]_{m',n'}}{[\mathbf{H}_\text{V}]_{m'-1,n'}}, \frac{[\mathbf{H}_\text{V}]_{m'-1,n'}}{[\mathbf{H}_\text{V}]_{m',n'}}\right)$      }
 \eIf{$|[\mathbf{H}_\text{V}]_{m',n+1}| > |[\mathbf{H}_\text{V}]_{m',n}|$}{
$n_l^\star = n' + \frac{1}{2} \min\left( \frac{[\mathbf{H}_\text{V}]_{m',n'}}{[\mathbf{H}_\text{V}]_{m',n'+1}}, \frac{[\mathbf{H}_\text{V}]_{m',n'+1}}{[\mathbf{H}_\text{V}]_{m',n'}}\right)$
        }{
$n_l^\star = n' - \frac{1}{2} \min\left( \frac{[\mathbf{H}_\text{V}]_{m',n'}}{[\mathbf{H}_\text{V}]_{m'-1,n'}}, \frac{[\mathbf{H}_\text{V}]_{m'-1,n'}}{[\mathbf{H}_\text{V}]_{m',n'}}\right)$      }
 }{
 \text{\textit{DOMP-MSLb Update:}}
\begin{align*}
&m_l^\star = m' +\frac{\tan(\frac{\pi}{M})}{\frac{\pi}{M}}\Delta_m; \quad n_l^\star =  n' +  \frac{\tan(\frac{\pi}{N})}{\frac{\pi}{N}} \Delta_n\\
&\text{\small $\Delta_m =  \mathbb{R}\left( \frac{[\mathbf{H}_\text{V}]_{m'-1,n'}  - [\mathbf{H}_\text{V}]_{m'+1,n'}}{2[\mathbf{H}_\text{V}]_{m',n'} - [\mathbf{H}_\text{V}]_{m'-1,n'} - [\mathbf{H}_\text{V}]_{m'+1,n'}} \right)$}\\
&\text{\small $\Delta_n = \mathbb{R}\left( \frac{[\mathbf{H_\text{V}}]_{m',n'-1}  - [\mathbf{H}_\text{V}]_{m',n'+1}}{2[\mathbf{H}_\text{V}]_{m',n'} - [\mathbf{H}_\text{V}]_{m',n'-1} - [\mathbf{H}_\text{V}]_{m',n'+1}} \right)$}
\end{align*}
}
\text{\scriptsize $\alpha_l^\star\hspace{-0.1cm} =\hspace{-0.1cm}[\mathbf{H}_\text{V}]_{m',{n'}}\hspace{-0.1cm} \left( \hspace{-0.1cm}\frac{\sin(\pi (m' - m_l^{\star}) )\sin(\pi (n' - n_l^{\star}))e^{-j\pi(m'-m_l^\star)}}{MN\sin{(\frac{\pi}{M} (m' - m_l^{\star}))}\sin{(\frac{\pi}{N}( n' - n_l^{\star}))}e^{-j\pi(n'-n_l^\star)}}\hspace{-0.1cm}\right)^{-1}$}
\\
\text{\scriptsize $[\mathbf{H}_{\text{V}_l}]_{m,n} \hspace{-0.1cm}=\hspace{-0.1cm} \frac{\alpha_l^\star}{{MN}}\frac{\sin(\pi (m - m_l^{\star}) )}{\sin{(\frac{\pi}{M} (m - m_l^{\star}))}}\frac{\sin(\pi (n - n_l^{\star}))}{\sin{(\frac{\pi}{N}( n - n_l^{\star}))}} \frac{e^{-j\pi(m-m_l^\star)}}{e^{-j\pi(n-n_l^\star)}}, \forall m, n $}\\
$\mathbf{y}_\text{res} = \mathbf{y}_\text{res} - \mathbf{A}\text{vec}(\mathbf{H}_{\text{V}_l})$\\
$l$ = $l$ + 1
}
}
\KwOut{$\mathbf{H}_\text{V} = \sum_{l=1}^L \mathbf{H}_{\text{V}_l}$}
\end{algorithm}

\begin{remark}
In some sense, what we describe in Algorithm 1 is a way to identify the true AoA-AoD for each MPC provided the DFT points of the MLb/SLb surrounding each of the Dirichlet kernels. The correct DFT peak points can often be obtained by the computationally tractable OMP projection strategy; steps 2 through 5 in Algorithm \ref{Algorithm:Dirichlet_Alg1}. Note that the steps of Algorithm 1 after the projection operation
(step 2) differ from standard OMP, which iterates between the projection and LS steps without exploiting the structure. 
Also, the steps for DOMP-MLb/MSLb are the same, except for the update steps enclosed in the \textit{if-else} statement (steps 6 through 20).
\end{remark}
\begin{remark}
The inherent disadvantage of the DOMP-MLb/MSLb is that Dirichlet kernels in the virtual domain must not overlap, implying that in the physical domain there cannot be closely spaced MPCs. In their current form, DOMP-MLb/MSLb are suitable only for scenarios such as terahertz communication \cite{gao2017fast} or a single MPC within each spatially separated cone \cite{2018VTC}. To overcome this limitation, we next propose a variant of the above method. 
\end{remark}

\subsection{DOMP-Local Optimization (DOMP-LO)}
The key idea here is to solve (\ref{Eq:Joint_Opt}) in an iterative fashion for each MPC over a localized space around nearby DFT peaks. If the point  $(m',n')$ where the DFT peaks for each MPC are provided, then the search space for the Dirichlet peak can be reduced to the range $([m'-1,m'+ 1],[n'-1,n'+ 1])$, which turns the joint optimization problem (\ref{Eq:Joint_Opt}) into a convex problem. The localized optimization problem (\ref{Eq:Joint_Opt}) can then be solved for the global optimum for each MPC separately; see step 4 in Algorithm~\ref{Algorithm:Dirichlet_Alg2}. Upon finding the peak location/strength of the Dirichlet kernel, it will be subtracted from the measurements $\mathbf{y}_\text{res}$ using step 5. This gets repeated until the stopping criterion is met. Finally, although DOMP-LO overcomes the drawbacks of DOMP-MLb/MSLb by solving the localized optimization problem (its ability to deal with multiple nearby MPCs), it requires more computation.

\begin{algorithm}[t!]
\KwIn{$\mathbf{y}$, $\mathbf{A}$, $\epsilon$ $\quad$\\ Initialization: $\mathbf{\mathcal{S}} = \{ \}$, $\mathbf{y}_\text{res} = \mathbf{y}$, ${ e } = || \mathbf{y}_\text{res}||_2$, $l$ = 1.}
\While{$e < \epsilon$}{
{
$j^{\star} = \arg \underset{j}{\max}| \mathbf{A}(:,j)^\text{T} \mathbf{y}_\text{res}|$\\
$m'$ = floor(${j^{\star}}/{M}$); \quad  $n'$ = mod(${j^{\star}},{N}$)\\
\text{Solve the localized version of (\ref{Eq:Joint_Opt}) :}
\begin{align*}
& \alpha_l^\star,\hat{m}_l^\star,\hat{n}_l^\star=  \underset{\alpha_l,{m}_l^\star,{n}_l^{\star}}{\text{min}}  ||\mathbf{y}_\text{res} - \mathbf{A} \text{vec}([\mathbf{H}_{\text{V}_l}]_{m',n'}) ||_2^2\\
&\textit{s.t.} \quad m \in (m'-1,m'+1), \quad n \in (n'-1,n'+1),\\ 
&\text{\scriptsize$\mathbf{H}_{\text{V}_l} =   \frac{\alpha_l}{{MN}}\frac{\sin(\pi (m - m_l^{\star}) )}{\sin{(\frac{\pi}{M} (m - m_l^{\star})}}\frac{\sin(\pi (n - n_l^{\star}))}{\sin{(\frac{\pi}{N}( n - n_l^{\star}))}}\frac{e^{-j\pi(n-n_l^\star)}}{e^{-j\pi(m-m_l^\star)}} $}.
\end{align*}
\\
$\mathbf{y}_\text{res} = \mathbf{y}_\text{res} - \mathbf{A} \text{vec}( [\mathbf{H}_{\text{V}_l}]) $\\
$l$ = $l$ + 1
}
}
\KwOut{$\mathbf{H}_\text{V} = \sum_{l=1}^L \mathbf{H}_{\text{V}_l}$}
\caption{Channel Estimation: DOMP-LO} \label{Algorithm:Dirichlet_Alg2}
\end{algorithm}

\section{Numerical Results}
In this section, the performance of the three proposed
channel estimation algorithms are provided, and compared to standard OMP. The performance is evaluated by the normalized mean square error 
(NMSE) = $\frac{||\mathbf{\hat{H}} - \mathbf{H}||_F^2}{||\mathbf{H}||_F^2}$. 
We consider a scenario with $M = N = G_\text{UE} = G_\text{BS} = 32$ and 3 MPCs. We deliberately place the MPCs away from the grid points. 
In particular, we place the AoA-AoD at a distance of  ($\pm0.1\zeta/2M,\pm0.1\zeta/2N$) from the middle of the randomly chosen adjacent grid points, where $\zeta$ is uniformly distributed in the range $[0,1]$. To avoid overlap in the virtual domain, the MPCs are spatially separated by at least $20^{\circ}$ from each other. This scenario helps evaluate the performance of DOMP-MLb/MSLb as noted in Remark 2. However, unlike the DOMP-MLb/MSLb, the DOMP-LO can be used even when the MPCs are closely spaced. All results presented below are averaged over 50 independent trials.  

\begin{figure}[t!]\vspace{-.3cm}
    \centerline{\includegraphics[scale=0.62]{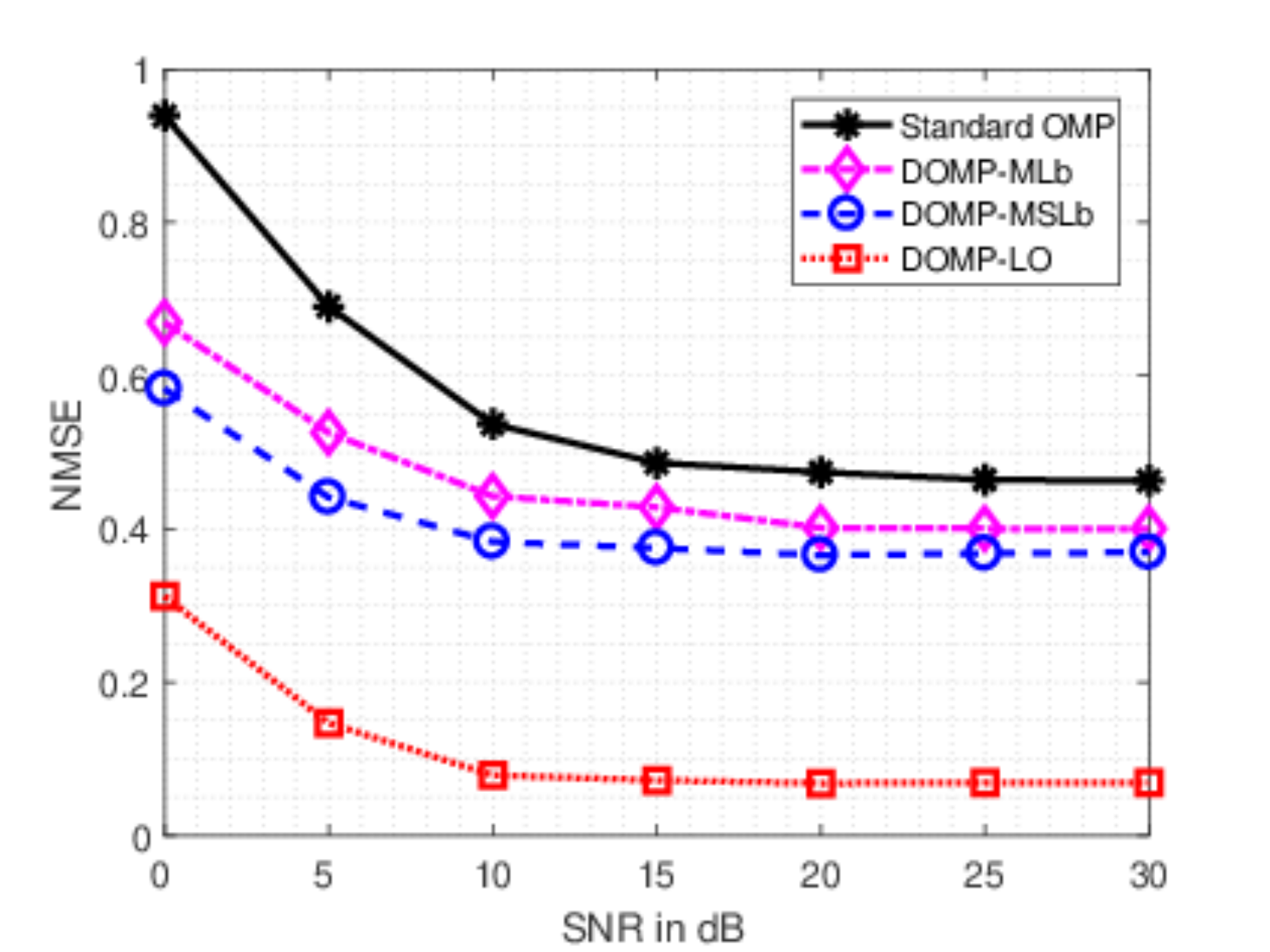} }\vspace{-1.1mm}
    \caption{Off-grid scenario with 3 MPCs: NMSE versus signal-to-noise ratio (SNR) in dB with measurements $M_\text{t}N_\text{t}$ = 100, where $M_{\rm t}$ and $N_{\rm t}$ are as in~\eqref{y_Equ}.}
    \label{fig:NMSE_SNR}\vspace{-.55cm}
\end{figure}

\begin{figure}[t!]
    \centerline{ \includegraphics[scale=0.62]{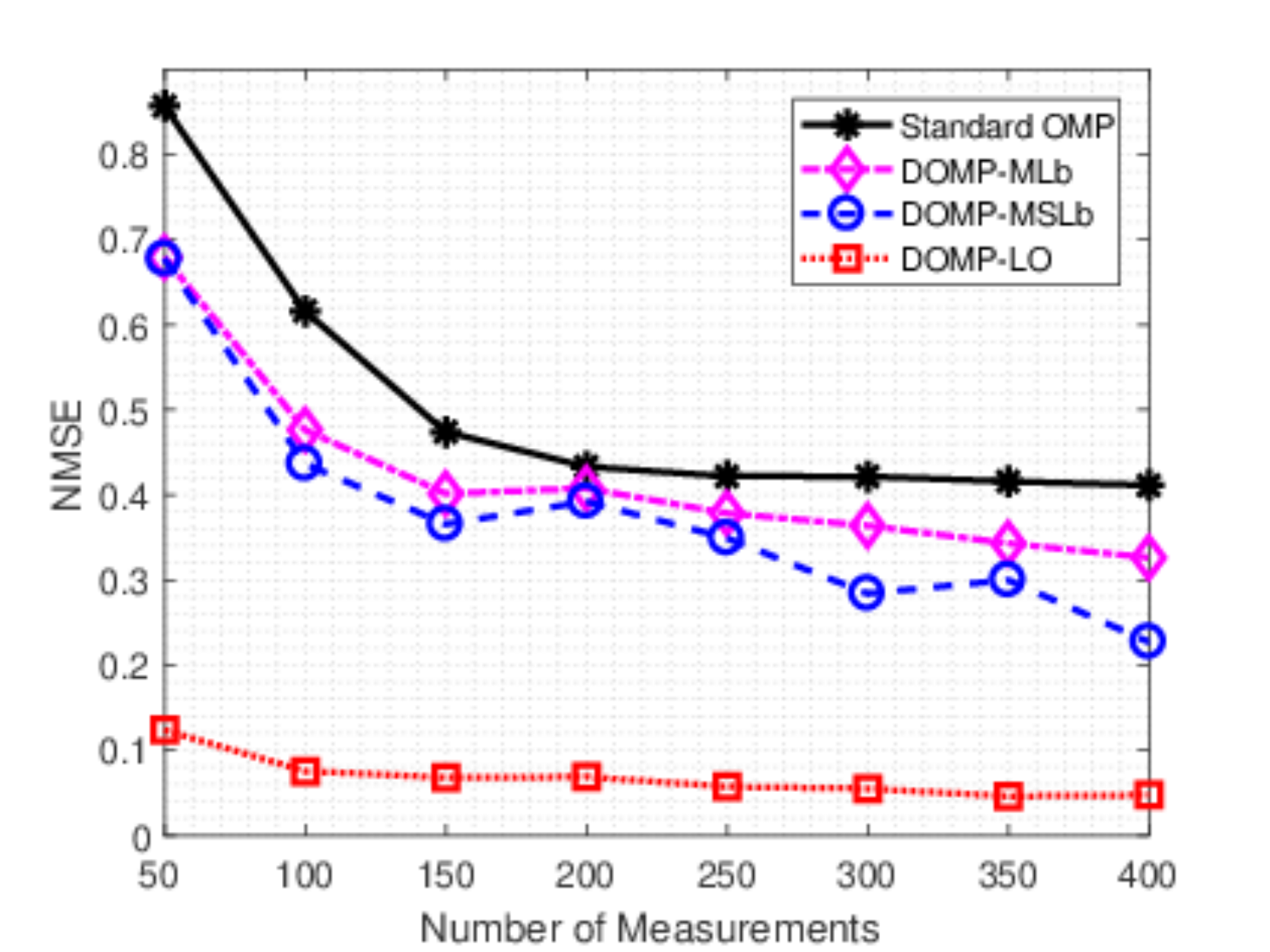}}\vspace{-1.1mm}
    \caption{Off-grid scenario with 3 MPCs: NMSE versus measurements ($M_\text{t}N_\text{t}$) with the SNR = 20 dB.}
    \label{fig:NMSE_Meas}
\end{figure}

Fig. \ref{fig:NMSE_SNR} and Fig. \ref{fig:NMSE_Meas} show how exploiting the Dirichlet kernel improves the channel estimation NMSE performance by accounting for off-grid effects, considering different SNRs and different number of measurements. The  performance improvement of the DOMP algorithms is mainly due to their ability to combat off-grid effects by traversing Dirichlet kernel peaks. 

Among the DOMP methods, the performance of DOMP-MSLb is slightly better than DOMP-MLb, as it is more robust than just considering the estimates from the MLb \cite{candan2011method}. The performance gap between standard OMP and DOMP-MLb (and MSLb) is significant, even in the low-SNR and small number of measurements regimes. Moreover, DOMP-LO (Algorithm~\ref{Algorithm:Dirichlet_Alg2}) outperforms other DOMP methods, although this requires more computation for the gradient and the Lagrange multipliers update in solving the localized problem. The detailed computational complexity analysis of the proposed methods is left for future work.
\section{Conclusion}\vspace{-0.1cm}
In this paper, we proposed low-complexity iterative OMP-based algorithms to exploit the implicit Dirichlet structure in the Fourier domain, and thus combat off-grid effects. The proposed DOMP-MLb/MSLb uses the MLb/SLb estimates to traverse to the Dirichlet peaks. On the other hand, the DOMP-LO method solves localized optimization problems to achieve a similar goal. Numerical results show that, in comparison to standard OMP our proposed algorithms achieve 
lower reconstruction errors in off-grid scenarios for a wide range of SNR and measurement levels.
\vspace{-0.1cm}
\bibliographystyle{IEEEtran}
\bibliography{IEEEabrv,Bib_File}
\end{document}